\documentclass[final,1p,times]{elsarticle}

\usepackage{amssymb}
\usepackage{bm}
\usepackage{epsfig}

\journal{Nuclear Physics A}

\begin{document}
\newcommand{\hszero}{h^{0s}_{\mbox{\tiny dNN}}}
\newcommand{\htzero}{h^{0t}_{\mbox{\tiny dNN}}}
\begin{frontmatter}



\title{Hadronic weak interaction in the two-nucleon system
with effective field theories}


\author[sa]{Shung-ichi Ando}
\author[ch]{Chang Ho Hyun}
\ead{hch@daegu.ac.kr}
\author[jw]{Jae Won Shin}

\address[sa]{Theoretical Physics Group, School of Physics and Astronomy, \\
The University of Manchester, Manchester, M13 9PL, UK}
\address[ch]{Department of Physics Education, Deagu University,
Gyeongsan 712-714, Korea}
\address[jw]{Department of Physics, Sungkyunkwan University,
Suwon 440-746, Korea}

\begin{abstract}
Weak interactions
in two-nucleon system 
at low energies are explored 
in the framework of effective field theory.
We review our resent 
calculations of parity-violating observables 
in radiative neutron capture on a proton at threshold
where both pionful and pionless theories are employed. 
%

\end{abstract}


\end{frontmatter}

\section{Introduction}

Exchanges of $\pi$, $\rho$ and $\omega$ mesons have long been
a standard way to describe the weak interactions of the baryon
at the hadronic level.
One of the important issues with the one-meson exchange (OME)
weak interaction is the value of the weak coupling constants.
A 
bench marking calculations for the weak coupling constants was
done with a quark model \cite{ddh80}.
Calculation of the weak coupling constants since then employed
various models and theories, and the results as a whole turned 
out to be consistent to the ''best values" in Ref.~\cite{ddh80}.
However, experimental determinations are yet quite controversial,
so 
a revealing true values of the weak coupling constants still 
remains as an open question. 

In the present situation, one can cast a question:
Is the OME picture sufficient to describe the hadronic weak dynamics?
It is quite recent that people started to consider the effective field 
theory (EFT), which already achieved good successes in the description 
of strong interactions and electromagnetic (EM) transitions in the low 
energy two-nucleon systems, in search for an answer to the question 
in the hadronic weak interaction problem~\cite{kssw,hyun01,savage01,
zhu05,liu07,hyun07,shin09}.
With the counting rules of EFT, we can expand the two-nucleon weak
potential systematically and perturbatively. 
Interestingly, EFT gives
terms that are absent in the OME potential, but are more important
than the heavy-meson terms in it.

In this work we consider parity-violating (PV) observables
in radiative capture of thermal neutron on a proton.
In the polarized neutron capture, it is known that the PV
asymmetry, $A_\gamma$, is dominated by the pion-exchange term 
in the weak OME potential. We thus employ an EFT that explicitly
includes pions while heavier degrees of freedom are integrated out,
and the short range dynamics are represented by two nucleon 
contact terms. 
In the unpolarized neutron capture, on the other hand, PV 
polarization, $P_\gamma$, is dependent on the $\rho$- and $\omega$-
exchange terms in the OME weak potential. 
Since heavy mesons are integrated out in the EFT, interactions
relevant to $P_\gamma$ are only the two-nucleon contact terms.
Therefore we are naturally lead to consider $P_\gamma$ with 
a theory where all the interactions are represented only with 
contact terms, so called the pionless theory.

In the next sections theories are briefly reviewed.
In the following section, we present the results and discussion, 
and in the last section we summarize the work.



\section{Theory}
\subsection{Pionful theory}
\begin{figure}
\begin{center}
\epsfig{file=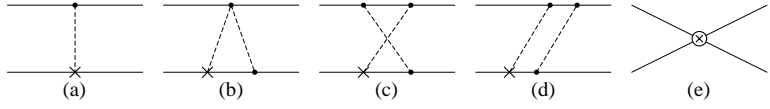, width=10cm}
\end{center}
\caption{Weak potential in the pionful theory at LO (OPE) and NNLO (TPE and
CT).
Solid lines denote nucleons, dashed lines do pions. Vertices with ``X'' 
denote PV interactions whereas those with a dot for PC ones. }
\label{fig:1}
\end{figure}

Order of a contribution (equivalently a Feynman diagram) 
is counted in powers of $Q/\Lambda_\chi$, 
where $Q$ is the momentum scale of the incoming 
and outgoing particles, and $\Lambda_\chi$ is the chiral 
symmetry breaking scale.
We count the order of $Q/\Lambda_\chi$ of a diagram with 
the counting rules of the heavy baryon chiral perturbation theory \cite{tsp93}.
Weak pion-nucleon Lagrangian is given by
\begin{equation}
{\cal L}_{\rm pv} = - \frac{1}{\sqrt{2}} h^1_\pi
\bar{N} \left( {\bm \tau} \times {\bm \pi} \right)^3 N,
\end{equation}
where $h^1_\pi$ is a weak pion-nucleon coupling constant.
At the leading order (LO) ($\propto Q^{-1}$), only the one-pion-exchange (OPE)
diagram contributes to the weak potential. 
There is no diagram at the order of $Q^0$, and the next-to-next leading order
(NNLO) consists of the two-pion-exchange (TPE) contributions and 
a two-nucleon contact term (CT) which has an unknown low energy constant (LEC).
Diagrams at LO and NNLO are shown in Fig.~\ref{fig:1}.
Weak potentials in Fig.~\ref{fig:1} in coordinate space can be written
in the form
\begin{eqnarray}
V_i(\bm{r}) = i (\bm{\tau}_1 \times \bm{\tau}_2)^z
(\bm{\sigma}_1 + \bm{\sigma}_2) \cdot \left[ \bm{p},\, v_i(r) \right],
\label{eq:general_form}
\end{eqnarray} 
where
${\bm p}$ is the momentum operator conjugate to the relative coordinate
${\bm r} = {\bm r}_1 - {\bm r}_2$.
In the Fourier transformation from momentum space to coordinate one,
we encounter ultraviolet divergence for the TPE and contact terms.
In order to regularize the divergence, we introduce 
the monopole 
cutoff function 
$\Lambda^2/(\Lambda^2 + q^2)$ for TPE and contact terms.
Thus the regularized OPE, TPE and CT terms $v^\Lambda_i(r)$ 
in Eq.~(\ref{eq:general_form}) read
\begin{eqnarray}
v^\Lambda_{1\pi}(r) &=& \frac{g_A h^1_\pi}{2 \sqrt{2} f_\pi}
 \frac{1}{4\pi r}
({\rm e}^{-m_\pi r} - {\rm e}^{-\Lambda r})\,, 
\label{eq:vr_ope}\\
v^\Lambda_{2\pi}(r) &=& \sqrt{2}\pi \frac{h^1_\pi}{\Lambda^3_\chi}
\left\{ g_A L^\Lambda (r)
- g^3_A \left[3 L^\Lambda (r) - H^\Lambda (r) \right]
\right\},
\label{eq:vr_tpe} \\
v^\Lambda_{\rm CT}(r) &=& - C^R_6 \Lambda^2 
\frac{{\rm e}^{-\Lambda r}}{4 \pi r}\,.
\label{eq:vr_ct}
\end{eqnarray}
Analytic 
expressions of the functions $L^\Lambda (r)$ and 
$H^\Lambda (r)$ can be found in Ref.~\cite{hyun07}, and
$C^R_6$ is the renormalized LEC in the contact term. 
The LEC $C^R_6$ can be determined model independently with experimental
data, but error bars of the data for the hadronic weak interaction
are too large to constrain the LEC with reasonable uncertainty.
In the theoretical calculation, $C^R_6$ is dependent on the regularization
schemes, and 
several choices are possible. 
In this work we consider the minimal
subtraction scheme, in which the renormalized LEC takes the form
\begin{equation}
C^R_6 = C_6(\mu) - h^1_\pi\frac{\pi g_A}{\sqrt{2} \Lambda^3_\chi}
(1 -3 g^2_A) \left[\frac{1}{\epsilon} -\Gamma + \ln(4 \pi) \right],
\end{equation}
where $C_6(\mu)$ is the bare LEC, a function of the renormalization 
point $\mu$ 
\cite{hyun07}, and $\Gamma = 0.5772$.
When the parity-odd potential is added to the parity-conserving (PC) 
potential, it causes parity mixture in the wave function,
and this parity admixture can contribute to 
EM transitions that are forbidden
in the pure parity states.

\subsection{Pionless theory with dibaryon fields}
With dibaryon fields in the pionless EFT, it has been shown in 
recent works \cite{ando05,ando06,ando08} that low energy observables
can be calculated with good convergence and accuracy.
Details for the strong and EM interactions in the pionless dibaryon
formalism can be found in those papers, and in this work we discuss
the PV terms in the theory.

At sufficiently low energies, pion mass can be treated as a large
scale. In this case, just as heavy mesons are integrated out in the pionful
theory, the contribution of pion exchange can be subsumed in the two-nucleon
contact terms. Then since we don't have 
mesons 
dynamically 
in the theory,
all the interactions are represented by the contact terms.
Effective weak Lagrangian for the low energy in the pionless EFT have 
been obtained in Ref.~\cite{zhu05}. 
Pionless Lagrangian with dibaryon fields can be obtained by replacing 
a two-nucleon field with a dibaryon one of the same quantum numbers.
We obtain Lagrangian for the PV dibaryon-nucleon-nucleon ($dNN$)
interaction as
\begin{eqnarray}
{\cal L}^0_{\mbox{\tiny PV}} &=&
\frac{h^{0 s}_{\mbox{\tiny dNN}}}{2 \sqrt{2\, \rho_d\, r_0}\, m_N^{5/2}} 
s^\dagger_3\, N^T 
\sigma_2 \sigma_i  \tau_2 \tau_3 
\frac{i}{2} \left(\stackrel{\leftarrow}\nabla - 
\stackrel{\rightarrow}\nabla \right)_i N +{\rm h.c.} \label{eq:Lwk0s}\\
& & + 
\frac{h^{0 t}_{\mbox{\tiny dNN}}}{2 \sqrt{2} \rho_d\, m_N^{5/2}} \,
t^\dagger_i\, N^T \sigma_2 \tau_2 
\frac{i}{2} \left(\stackrel{\leftarrow}\nabla - 
\stackrel{\rightarrow}\nabla \right)_i N
+{\rm h.c.}, \label{eq:Lwk0t}
\end{eqnarray}
where $\hszero$ and $\htzero$ denote weak $dNN$ coupling constants 
for parity mixing for the $^1 S_0$ and $^3 S_1$ dibaryon
states, respectively.
Spin-isospin operator $\sigma_2 \sigma_i \tau_2 \tau_a$ in
Eq.~(\ref{eq:Lwk0s}) projects two-nucleon system to $^3 P_0$
state. PV vertex given by Eq.~(\ref{eq:Lwk0s}) therefore mixes
$^3 P_0$ in the $^1 S_0$ state.
Similarly, $\sigma_2 \tau_2$ in Eq.~(\ref{eq:Lwk0t}) is 
projection operator to $^1 P_1$ state, and thus the Lagrangian
mixes $^1 P_1$ state in the $^3 S_1$ state.
We note that the terms in Eqs.~(\ref{eq:Lwk0s},\ref{eq:Lwk0t}) are 
relevant to the PV polarization in unpolarized neutron capture.
They are part of the general Lagrangian that accounts for all the possible
parity-mixing modes \cite{dan09}.

\section{Result and Discussion}
\subsection{Asymmetry in polarized neutron capture}
Parity-violating asymmetry $A_\gamma$ in $\vec{n} p \to d\gamma$
is defined as
\begin{eqnarray}
\frac{d \sigma}{d \Omega} \propto 1 + A_\gamma \cos \theta,
\end{eqnarray}
where $\theta$ is an angle of the photon momentum with respect to the
neutron polarization.
We employ a hybrid scheme in the calculation, 
where only weak potentials are obtained from EFT, while strong 
potential is described with a phenomenological model, 
Argonne v18 and 
the EM operators are assumed to satisfy Siegert's theorem.
Weak pion-nucleon coupling constant $h^1_\pi$ is of the order of
$10^{-7}$ so it is sufficient to work at its linear order. 
PV asymmetry is then proportional to $h^1_\pi$, and thus we write
the result as
\begin{eqnarray}
A_\gamma = a_\gamma h^1_\pi.
\end{eqnarray}
The coefficient $a_\gamma$ depends on various inputs such as
strong interaction, weak interaction, EM operators, cutoff value. 
The results in Table~\ref{tab:result1} show the contribution to $a_\gamma$
from OPE and TPE terms, and their dependence on the cutoff $\Lambda$ and 
renormalization scale $\mu$.
\begin{table}
\begin{center}
\begin{tabular}{lcrrrr}
\hline
$\Lambda$ (MeV) & & 500 & 1000 & 1500 & 2000 \\
\hline
$a_\gamma \mbox{(OPE)}$ & &
$-0.0992$ & $-0.1104$ & $-0.1117$ & $-0.1119$ \\
\hline
 & $\mu = 2 m_\pi$ & $-0.0987$ & $-0.0969$ & $-0.0966$ & $-0.0966$ \\
$a_\gamma \mbox{(total)}$ & $\mu = m_\rho$ &
$-0.0885$ & $-0.0917$ & $-0.0941$ & $-0.0952$ \\
 & $\mu = 1\, {\rm GeV}$ & $-0.0859$ & $-0.0904$ & $-0.0935$ & $-0.0949$ \\
\noalign{\smallskip}\hline
\end{tabular}
\end{center}
\caption{Contribution from LO is denoted by OPE, and those the NNLO contribution
can be obtained by subtracting OPE from the total.}
\label{tab:result1}
\end{table}
Dependence on the cutoff $\Lambda$ is 
non-negligible for both OPE and total value (OPE and TPE), 
and it is
more drastic for the NNLO (total subtracted by OPE). 
Difference is relatively large between $\Lambda = 500$ MeV and 1000 MeV,
i.e. about 11\% increase in magnitude for LO and about 54\% for NNLO.
However, the values become quite stable when $\Lambda$ is more than 1000 MeV,
and the change reduces to about 1\% for LO and 18\% for NNLO.
Dependence on the renormalization scale $\mu$ is already sizable, and
the largest gap is about 15\% of the value of the smallest magnitude, $-0.0859$.
However, the cutoff dependence almost disappears when the renormalization 
scale is two times the pion mass, $\mu=2m_\pi$.
These values can be regarded as fixed-point results up to NNLO.

\subsection{Polarization in unpolarized neutron capture}
\begin{figure}[tbp]
\begin{center}
\epsfig{file=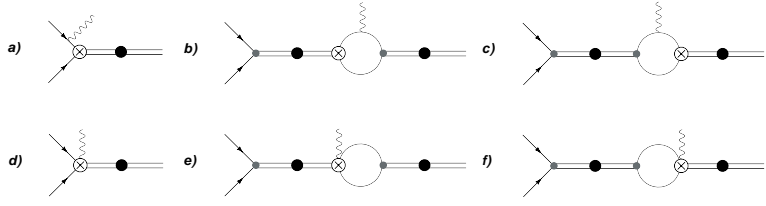, width=10cm}
\end{center}
\caption{PV diagrams for $np$ capture at LO.
Single solid line denotes a nucleon, wavy line a photon, and
a double line with a filled circle stands for dressed dibaryon propagator.
Circle with a cross represents a PV $dNN$ vertex.}
\label{fig:2}
\end{figure}

Parity-violating polarization $P_\gamma$ in $np \to d\gamma$ is defined as
\begin{equation}
P_\gamma = \frac{\sigma_+ - \sigma_-}{\sigma_+ + \sigma_-},
\end{equation}
where $\sigma_+$ and $\sigma_-$ are the total cross section for the photons
with right and left helicity, respectively. 
While we calculate $A_\gamma$ in $\vec{n}p \to d\gamma$ in a hybrid way,
$P_\gamma$ is calculated self-consistently in the pionless dibaryon
formalism. Both PC and PV amplitudes are calculated with the same counting 
rules, and the PC and PV diagrams are truncated at the same order.
For the PC amplitude for $np \to d\gamma$, we employ the result in 
Ref.~\cite{ando05},
\begin{eqnarray}
Y &=& \frac{\sqrt{2 \pi}}{m^2_N} \sqrt{\frac{\gamma}{1 -\gamma \rho_d}}
\left[ (1 + \kappa_V) ( 1 - \gamma a_s) - \gamma^2 a_s L_1 \right].
\end{eqnarray}
Feynman diagrams that 
contribute to the PV amplitude at LO are shown in Fig.~\ref{fig:2}.
Taking the zero momentum approximation, we obtain the PV amplitude at LO as
\begin{eqnarray}
Z = \frac{1}{m^2_N \sqrt{m_N \rho_d}} 
\sqrt{\frac{\gamma}{1-\gamma\rho_d}} 
\left[\htzero \left( \frac{1}{2} - \frac{1}{3} \gamma a_s \right)
+ \frac{1}{6} \hszero \gamma a_s \right],
\label{eq:pvz}
\end{eqnarray}
and the PV polarization $P_\gamma$ at LO reads
\begin{eqnarray}
P_\gamma &=&  - 2 \frac{{\rm Re} (Y Z^*)}{|Y|^2}
\nonumber \\ &=&
- \sqrt{\frac{2}{\pi m_N \rho_d}} 
\frac{\left(\frac{1}{2}-\frac{1}{3}\gamma a_s\right) \htzero 
+ \frac{1}{6} \gamma a_s \hszero }
{(1+\kappa_V)(1-\gamma a_s)-\gamma^2 a_s L_1}. 
\label{eq:pgamma}
\end{eqnarray}
Since the parameters other than $\htzero$ and $\hszero$ are well fixed,
measurements of $P_\gamma$ provides a relation that can constraint the
values of $\htzero$ and $\hszero$.

\section{Summary}
We considered weak interaction at the hadron level with effective field
theories. 
With the pionful theory, we derived weak two-nucleon potential
up to NNLO. Parity-violating asymmetry $A_\gamma$ 
calculated with the weak potential
shows that the perturbative expansion converges reasonably, and
the results are non negligibly dependent on the cutoff value and 
renormalization scale. However, the LO contribution is sufficiently 
dominant that we can determine the first significant digit of the 
weak pion-nucleon coupling
constant $h^1_\pi$ if $A_\gamma$ is measured precisely.

With the pionless theory, we calculated the parity-violating
polarization $P_\gamma$ in $np \to d\gamma$.
In the hybrid calculation adopted in the calculation of $A_\gamma$,
orders of the strong interaction and EM operators are not countable
because they are quoted from the formulations where perturbative
expansion is absent. Therefore, order mismatch could be unavoidable 
in the hybrid calculation.
On the other hand, $P_\gamma$ is calculated in a single theory,
and the order of the diagrams is fixed to a single order.
The result is dependent on the weak coupling constants for
$^1 S_0 - ^3 P_0$ mixing and $^3 S_1 - ^1 P_1$ one.
Measurement of $P_\gamma$ is expected to provide a constraint to
reduce the possible space for these weak coupling constants.

\section*{Acknowledgments}
This research was supported by the Daegu University Research Grant, 2008.

\end{document}